\documentclass[letterpaper, 10 pt, conference]{ieeeconf}  % Comment this line out if you need a4paper

\IEEEoverridecommandlockouts                

\usepackage{graphicx,dblfloatfix} % for pdf, bitmapped graphics files
\usepackage{epsfig} % for postscript graphics files
\usepackage{mathptmx} % assumes new font selection scheme installed
\usepackage{times} % assumes new font selection scheme installed
\usepackage{amsmath} % assumes amsmath package installed
\usepackage{amssymb}  % assumes amsmath package installed
\usepackage{cite}
 \usepackage{tikz}
 \usepackage{graphicx,color,psfrag}
\usetikzlibrary{shapes,arrows,shadows,calc,patterns,decorations.pathmorphing,decorations.markings}
\usepackage{verbatim}
\definecolor{tublue}{RGB}{0,166,214}
\definecolor{tuyellow}{RGB}{247,235,144}
\definecolor{blue}{RGB}{0,166,214}
 \definecolor{lblue}{RGB}{119,192,215}
 \definecolor{lred}{RGB}{236,127,44}
 \newtheorem{proposition}{Proposition}
 \newtheorem{remark}{Remark}

\title{Convex Model Predictive Control for Down-regulation  Strategies in Wind Turbines}

\author{Jean Gonzalez Silva,  Riccardo Ferrari and Jan-Willem van Wingerden$^{*}$ 
\thanks{$^{*}$ Delft University of Technology, Delft, 2628CD The Netherlands {\tt\small \{J.GonzalezSilva, R.Ferrari, J.W.vanWingerden\}@tudelft.nl}}%
}

\begin{document}

\maketitle
\thispagestyle{empty}
\pagestyle{empty}

%%%%%%%%%%%%%%%%%%%%%%%%%%%%%%%%%%%%%%%%%%%%%%%%%%%%%%%%%%%%%%%%%%%%%%%%%%%%%%%%
\begin{abstract}
% Wind turbines are often set for maximum power extraction while protecting their components. Fulfilling the challenges in wind energy, turbine down-regulation allows to balance power generation and grid loads, as well as to provide auxiliary grid services, such as frequency regulation. Although wind availability is limited depending on the weather conditions and grid demand, the quality of wind energy can be improved by making use of the kinetic energy of the turbine's rotor. This paper shows the degree of freedom in the kinetic energy in down-regulated turbines where different down-regulation strategies are explored using convex model predictive control at a single wind turbine. Also a constraint regarding flow stability is derived. The results show the performance of the developed down-regulation strategies. By achieving high rotor speeds, the turbine benefits from keeping power tracking when power saturation occurs. In the study case, it reaches about 10 times longer, however its operation becomes closer to unstable flow regions also the corresponding aerodynamic load increases by about 18\%.

Wind turbine (WT) controllers are often geared towards maximum power extraction, while suitable operating constraints should be guaranteed such that WT components are protected from failures. Control strategies can be also devised to reduce the generated power, for instance to track a power reference provided by the grid operator. They are called down-regulation strategies and allow to balance power generation and grid loads, as well as to provide ancillary grid services, such as frequency regulation. Although this balance is limited by the wind availability and grid demand, the quality of wind energy can be improved by introducing down-regulation strategies that make use of the kinetic energy of the turbine dynamics. This paper shows how the kinetic energy in the rotating components of turbines can be used as an additional degree-of-freedom by different down-regulation strategies. In particular we explore the power tracking problem based on convex model predictive control (MPC) at a single wind turbine. The use of MPC allows us to introduce a further constraint that guarantees flow stability and avoids stall conditions. Simulation results are used to illustrate the performance of the developed down-regulation strategies. Notably, by maximizing rotor speeds, and thus kinetic energy, the turbine can still temporarily guarantee tracking of a given power reference even when occasional saturation of the available wind power occurs. In the study case we proved that our approach can guarantee power tracking in saturated conditions for 10 times longer than with traditional down-regulation strategies.

\end{abstract}

%%%%%%%%%%%%%%%%%%%%%%%%%%%%%%%%%%%%%%%%%%%%%%%%%%%%%%%%%%%%%%%%%%%%%%%%%%%%%%%%
\section{INTRODUCTION}

%\emph{- Problem definition}
% In the transition to renewable energy sources, several countries have exceed the penetration level of renewable generation in 15\% of their overall power-generation mix, which many of them (e.g., Spain, Portugal, Ireland, Germany, Denmark and the United States) have already experienced instantaneous penetration levels higher than 50\% \cite{IEA-RETD}. Due to the significant penetration levels of wind power in the electricity consumption, grid operators increase their demand for system services to be provided by wind turbines (WTs).
\noindent In the transition to renewable energy sources, several countries reached a penetration level of renewable generation of more than 15\% of their overall power-generation mix. Many of them (e.g., Spain, Portugal, Ireland, Germany, Denmark and the United States) have already crossed this threshold significantly, and experienced instantaneous penetration levels higher than 50\% \cite{IEA-RETD}. Due to such significant contribution of renewable energy sources, including wind power, grid operators are increasing their demand for ancillary services to be provided by wind turbines (WTs).

%Wind power systems fall under the category of variable generation, as the maximum available power varies over time, and it cannot  be predicted with perfect accuracy.

% In active power control (APC), the WTs respond to grid requirements through control of their power output \cite{fleming2016}. 
% At a single turbine level, a power command signal provided by the system operator can be followed in order to balance power supply with a required load. As WTs are essentially non-linear systems, there are several down-regulation methodologies that achieve power tracking \cite{vanderhoek2018, lio2018, juangarcia2018}. Based on them, in this work we elucidate the degree of freedom in terms of the kinetic energy associated with the rotating parts, including blades and generator.
In particular, grid operators can make use of so-called \emph{Active Power Control} (APC) to request turbines to provide a given reference power output \cite{fleming2016}. The power reference command sent to all generators will guarantee that, at grid level, supply and demand are balanced and grid frequency is stabilized.
As the power that a WT can generate is upper bounded by the available power in the incoming wind flow, WTs can only be \emph{down-regulated}, that is operate in a way to track a power reference that is lower than the theoretical available maximum. Due to the nonlinearities present in the dynamics of WTs, several down-regulation methodologies that achieve power tracking are possible~\cite{vanderhoek2018, lio2018, juangarcia2018}. 
%Based on them, in this work we elucidate the degree of freedom in terms of the kinetic energy associated with the rotating parts, including blades and generator.

%

%As the rotating parts can compromise the safe integrity of the system it should be also taken into consideration, in which down-regulation responses can have a negative effect on structural loading. Nevertheless, the flow stability along the blades should be kept avoiding blade stall and power oscillation at given WTs' operation \cite{}.

%Future systems with high degrees of non-synchronous penetration can achieve power system stability by wind turbine controllers providing the expected grid services. 

Still, existing down-regulation strategies were developed for steady state conditions only, and cannot directly take into account available information on changing wind conditions, such as those provided by short time weather forecasts or LIDAR measurements. Furthermore, they do not accommodate directly the need to minimize structural loads on the WT, which on the long period can lead to premature failures.

In order to address these challenges, in this paper we propose a down-regulation approach based on convex \emph{Model Predictive Control} (MPC). We will show how all the major down-regulation strategies present in the literature can be implemented with the proposed MPC approach. Furthermore, we will introduce a novel down-regulation strategy based on the maximization of kinetic energy, and show its benefits in guaranteeing power tracking also during occasional periods of saturation, when the reference power from the grid is larger than the available power in the wind flow.

MPCs approaches have already demonstrated their potential in several works on wind turbine and wind farm \cite{riverso2016model} control. An MPC formulation based on power flow and energy was presented in~\cite{hovgaard2013}, while~\cite{shaltout2020} and~\cite{jain2020} have extended it by including the tower flexural model and by considering the presence of faults, respectively. By assuming the knowledge of future demanded power and wind variations, in~\cite{tielens2017} the authors are able to damp grid frequency oscillations by storing and releasing the WT kinetic energy. The use of kinetic energy as an energy reserve for grid stabilization is also explored in~\cite{odgaard2016}, where power generation can be increased by temporarily supplying kinetic energy from the rotor. To the best of the authors' knowledge, anyway, no work did consider the problem of promoting flow stability on the WT blades during down regulation. Indeed operating in low flow stability regions can lead to rotor speed oscillations, undesirable turbine responses and ultimately cause stall conditions~\cite{lio2018}.

% In this work, we analyse the performance of down-regulation methods with a convex model predictive control for the NREL 5 MW WT \cite{jonkman2009}. The wind turbine dynamics is simulated using NREL's Fatigue, Aerodynamics, Structures and Turbulence (FAST) tool \cite{jonkman2005}. Although there is a degree of freedom in the kinetic energy of down-regulated turbines, the flow stability should be kept to avoid oscillations and undesirable turbine responses leaded by stall conditions \cite{lio2018}.
% Different down-regulation methods are evaluated by comparing their associated operation input commands and responses, such as the aerodynamic loads. As a result of distinct kinetic energies corresponding to the used different methods, the power tracking could be temporally kept longer in situations of extreme grid demand or low power availability in the wind.  %Further, the WT performance is analyzed as well as the grid’s perspective.

The contribution of the present paper is three fold.
\begin{itemize}
    \item We develop a general convex MPC framework for power tracking on wind turbines which includes the kinetic energy as a degree of freedom;
    \item We extend the cost function in order to minimize aerodynamic loads, and add a constraint that enforces flow stability;
    \item We present a simulation study based on the NREL 5 MW WT~\cite{jonkman2009} and compare different down-regulation strategies under saturated conditions in OpenFAST~\cite{openfast}.
\end{itemize}

A key ingredient for obtaining a convex MPC formulation in the present case is to use a linear model of the WT dynamics, expressed in energy form. Such form allows to remove the non-linear relationship among rotor speed, blade pitch angles, and wind speed from the optimization problem. The aerodynamic rotor power is then chosen as an optimisation variable, which is constrained by a piecewise affine approximation of the available wind power.
This formulation allows naturally to include the kinetic energy as a degree of freedom and leads to a linear optimization problem. Due to this freedom, an extra objective can be added to the optimization problem and thus recover the different existing down-regulation strategies. Finally, we show how to avoid stall conditions by implementing a further linearized constraint .
% Each down-regulation strategy has a particular behavior whenever turbine saturation occurs, i.e. the wind available power is less than the one demanded by the grid.
As will be seen in the simulation results,
the time period during which the demanded power is tracked, by maximizing kinetic energy, can be up to ten times longer with respect to other down regulation strategies.

The structure of this paper is as follows. First, the wind turbine model is decribed in Section \ref{wtmodel}. Next, down-regulation strategies are analyzed in terms of kinetic energy and flow stability in Section \ref{dr}.
Section \ref{CMPC} introduces our proposed convex MPC for different down-regulation methods and their constraints. A simulation study under saturated conditions is presented in Section \ref{simul}. Finally, the paper is concluded in Section \ref{conc} .

\section{Wind Turbine Model} \label{wtmodel}
The non-linear wind turbine dynamics can be modelled using the rotor torque balance equation. By considering a rigid shaft and neglecting losses, this leads to the following model.%, being the shaft angle displacement negligible.
\begin{equation}\label{basicmodel}
\dot{\omega}_\mathrm{g}(t) = \frac{1}{J} \left[ \frac{1}{G_\mathrm{B}} T_\mathrm{r}(t) - T_\mathrm{g}(t) \right],
\end{equation}
where $J$ is the equivalent moment of inertia of the rotor-generator-drive-train assembly referred to the high speed shaft, $\dot{\omega}_\mathrm{g}(t)$ the generator acceleration, $T_\mathrm{r}(t)$ the aerodynamic torque, $G_\mathrm{B}$ the gearbox ratio and $T_\mathrm{g}(t)$ the generator torque.

The non-linearity due to the aerodynamic torque relation can be expressed as 
\begin{subequations}
	\begin{align}
	T_\mathrm{r}(t) = & \dfrac{0.5}{\omega_\mathrm{r}(t)} \rho A_\mathrm{r} v^3(t) C_\mathrm{P}(\lambda(t), \theta(t)) \label{taero} \\
	= & \Phi (v(t), \omega_\mathrm{r}(t), \theta(t))v^3(t)/\omega_\mathrm{r}(t) \label{taerob} \\
    = & 0.5 \rho A_\mathrm{r} R v^2(t) C_\mathrm{Q}(\lambda(t), \theta(t)), \label{taeroc} 
    \end{align}
\end{subequations}
where $\rho$ the air density, $A_\mathrm{r}$ the rotor area, $\theta(t)$ the collective blade-pitch angle, and $\lambda(t) = R\omega_\mathrm{r}(t)/v(t)$ is the tip-speed ratio, being $R$ the rotor radios and $\omega_\mathrm{r}(t)=\omega_\mathrm{g}(t)/G_\mathrm{B}$ the rotor speed.
The representation in \eqref{taero} is as function of the power coefficient $C_\mathrm{P}(\lambda(t), \theta(t))$.
In \eqref{taerob}, instead the function $\Phi(v(t), \omega_\mathrm{r}(t), \theta(t))=0.5 \rho A_\mathrm{r} C_\mathrm{P}(\lambda(t), \theta(t))$ is defined. Finally, the rotor torque in \eqref{taeroc} is a function of the torque coefficient $C_\mathrm{Q}(\lambda(t), \theta(t))$, where it holds $C_\mathrm{Q}(\lambda(t), \theta(t))=C_\mathrm{P}(\lambda(t), \theta(t))/\lambda(t)$.
%as
%\begin{equation}
%\Phi (v(t), \omega_\mathrm{r}(t), \theta(t)) = 0.5 \rho A_\mathrm{r} C_\mathrm{P} \left( \dfrac{R\omega_\mathrm{r}(t)}{v(t)}, \theta(t) \right).
%\end{equation}

 The collective blade pitch angle, generator speed and torque are limited by their upper and lower bounds as follows:
\begin{subequations}\label{const1}
	\begin{align}
	\theta_\mathrm{min} \leq \theta(t) \leq \theta_\mathrm{max}; \label{pitchconst} \\
	\omega_\mathrm{g, min} \leq \omega_\mathrm{g} (t) \leq \omega_\mathrm{g, max}; \label{rotorconst} \\
	0 \leq T_\mathrm{g}(t) \leq T_\mathrm{g, max} \label{torqueconst}.
	\end{align}
\end{subequations}

The aerodynamic power extracted from the wind by the rotor is given as
\begin{equation} \label{rotorpower}
P_\mathrm{r}(t) = T_\mathrm{r}(t) \omega_\mathrm{r}(t) = 0.5 \rho A_\mathrm{r} v^3(t) C_\mathrm{P} (\lambda (t), \theta(t)),
\end{equation}
while the electrical generator power is given by
\begin{equation}\label{powergen}
P_\mathrm{g} (t)= \eta_\mathrm{g} T_\mathrm{g}(t) \omega_\mathrm{g}(t),
\end{equation}
where $\eta_\mathrm{g}$ is the generator efficiency. 

The electrical generator power is constrained by
\begin{equation}
0 \leq P_\mathrm{g}(t) \leq P_\mathrm{g, rated},
\end{equation}
where $P_\mathrm{g, rated}$ is the rated generator power.

In terms of power flow and energy, the dynamics in \eqref{basicmodel} is rewritten as
\begin{equation} \label{model}
\dot{K}(t)=P_\mathrm{r}(t) - \frac{1}{\eta_\mathrm{g}}P_\mathrm{g}(t),
\end{equation}
where $K(t)$ is the kinetic energy stored in the rotating components and relates to the generator speed as
\begin{equation}\label{kinetic_energy}
    K(t) = \frac{J \omega_\mathrm{g}^2(t)}{2}.
\end{equation}
%  In the proposed convex MPC, the variables $K$, $P_\mathrm{r}$ and $P_\mathrm{g}$ are considered as optimization variables, and \eqref{model} is the dynamic equation in the MPC problem.
 In the proposed convex MPC formulation, $P_\mathrm{r}$ and $P_\mathrm{g}$ in \eqref{model} are chosen to be the decision variables.
The set of constraints in \eqref{const1} needs then to be rewritten as a function of $P_\mathrm{r}$ and $P_\mathrm{g}$ and $K$ as well, as the latter depends on the former via the system dynamics \eqref{model}. Note that the transformed constraints should be also convex to lead to a convex problem \cite{boyd2004}. Using \eqref{kinetic_energy} and \eqref{powergen}, the rotor speed and generator torque constraints from \eqref{rotorconst} and \eqref{torqueconst} can be expressed respectively as 
\begin{equation} \label{constLinear1}
(J/2) \omega_\mathrm{g, min}^2 \leq K(t) \leq (J/2) \omega_\mathrm{g, max}^2,
\end{equation}
\begin{equation} \label{constLinear2}
0 \leq P_\mathrm{g}(t) \leq \eta_\mathrm{g} \sqrt{(2/J)K(t)} T_\mathrm{g, max}.
\end{equation}
Since $\sqrt{(2/J)K(t)}$ is a concave function, \eqref{constLinear2} is a convex constraint on $P_\mathrm{g}(t)$ and $K(t)$.
Defining the available power as
$$ P_\mathrm{av}(v, K) = \max_{\theta_\mathrm{min}\leq \theta \leq \theta_\mathrm{max}} \Phi(v, (1/G_\mathrm{B})\sqrt{(2/J)K},\theta) v^3, $$
the aerodynamic rotor power constraint is set as
\begin{equation} \label{constLinear3}
P_\mathrm{r}(t) \leq P_\mathrm{av}(v(t), K(t)).
\end{equation}
This includes the constraint \eqref{pitchconst} in the formulation.

For a range of wind speeds and blade pitch angles and realistic $\Phi$ functions, the available power turns out to be a concave function of $K$. Therefore, by fitting $k$ piecewise affine functions~\cite{magnani2009}, the available power can be approximated as 
$$ \hat{P}_{\mathrm{av}, v_\mathrm{i}} (K(t)) = min \{ a_\mathrm{1} K(t) + b_\mathrm{1} , \, ... \, , a_\mathrm{k} K(t) + b_\mathrm{k} \} v_i^3, $$
where a linear interpolation is done between different wind speeds to obtain $\hat{P}_\mathrm{av}(v(t),K(t))$.  
\begin{equation}
\hat{P}_\mathrm{av}(v(t),K(t))=
(1-\Theta) \hat{P}_\mathrm{av, v_\mathrm{1}} (K(t)) + \Theta \hat{P}_\mathrm{av, v_\mathrm{2}} (K(t)),
\end{equation}
with $\Theta=\dfrac{v(t) - v_1}{v_2 - v_1}$. The linear interpolation of concave functions results in a concave function.

The thrust force, which
presents another non-linear behavior, is modelled as
\begin{equation}\label{thrust}
F_\mathrm{T}(t) = 0.5 \rho A_\mathrm{r} v^2(t) C_\mathrm{T}(\lambda(t), \theta(t)),
\end{equation}
where $C_\mathrm{T}$ is the thrust coefficient. 

In order to develop the down-regulation strategy that minimizes thrust force, we approximate the thrust force through a linearization with respect to the aerodynamic rotor power and kinetic energy by assuming the knowledge of the wind speed at each time-step as follows.

First, the power and thrust coefficients are expressed with a first-order Taylor series expression around the current kinetic energy $K^*$ and blade pitch angle $\theta^*$, 
\begin{equation}\label{lcp} 
\begin{split}
	C_\mathrm{P}(K(t), \theta(t)) \approx  	C_\mathrm{P}(K^*, \theta^*) 
	+ 
	\left. \dfrac{\partial C_\mathrm{P}}{\partial K} \right|_{K^*,\theta^*} (K(t) - K^*)
	\\
	+ \left. \dfrac{\partial C_\mathrm{P}}{\partial \theta} \right|_{K^*,\theta^*} (\theta(t) - \theta^*)  =
	q_\mathrm{P}(t) \theta(t) + r_\mathrm{P}(t) K(t) + s_\mathrm{P}(t),
\end{split}
\end{equation}
\begin{equation}\label{lct} 
\begin{split}
	C_\mathrm{T}(K(t),\theta(t)) \approx C_\mathrm{T}(K^*, \theta^*) 
	+ 
	\left. \dfrac{\partial C_\mathrm{T}}{\partial K} \right|_{K^*,\theta^*} (K(t) - K^*)
	\\
	+ \left. \dfrac{\partial C_\mathrm{T}}{\partial \theta} \right|_{K^*,\theta^*} (\theta(t) - \theta^*)
	 = q_\mathrm{T}(t) \theta(t) + r_\mathrm{T}(t) K(t) + s_\mathrm{T}(t),
\end{split}
\end{equation}
where $q_\mathrm{P}(t)$, $r_\mathrm{P}(t)$, $s_\mathrm{P}(t)$, $q_\mathrm{T}(t)$, $r_\mathrm{T}(t)$ and $s_\mathrm{T}(t)$ are the corresponding time-varying parameters.

Then, combining \eqref{lcp} with \eqref{rotorpower} 
and  \eqref{lct} with \eqref{thrust}, 
and eliminating the collective blade pitch angle, an affine relationship can be derived at each time-step as

\begin{equation}\label{thrustlinear}
\hat{F}_\mathrm{T}(t) =
 Q_\mathrm{F_T}(t) P_\mathrm{r}(t) + R_\mathrm{F_T}(t) K(t) + S_\mathrm{F_T}(t)
\end{equation}
with $Q_\mathrm{F_T}(t) = \left( \dfrac{q_\mathrm{T}(t)}{q_\mathrm{P}(t) v(t)} \right)$, $R_\mathrm{F_T}(t) = 0.5 \rho A_r v(t)^2 \left( r_\mathrm{T}(t) - r_\mathrm{P}(t) \dfrac{q_\mathrm{T}(t)}{q_\mathrm{P}(t)} \right)$, and $S_\mathrm{F_T}(t)= 0.5 \rho A_r v(t)^2 \left( s_\mathrm{T}(t) - s_\mathrm{P}(t) \dfrac{q_\mathrm{T}(t)}{q_\mathrm{P}(t)} \right)$.

\section{DOWN-REGULATION} \label{dr}

%when the turbine is not required to extract the maximum power from the wind. 
%This can lead to a better operation of the turbines considering them as assets by reducing their usage.  
As discussed earlier, there are several practical benefits in being able to down-regulate turbines to track a specific demanded power from the grid. 
% This brings benefits for the power plant, such as in terms of structural stresses compared to maximum energy extraction and supporting auxiliary grid services. % by  avoiding excessive aerodynamic loading. %and when no capacitance storage management system are available on the grid. 
However, there are multiple control solutions to down-regulate wind turbines. In Table \ref{tablestrategies}, the main strategies from the literature are summarized and qualitatively compared in terms of their capabilities to track a power reference, reduce structural loads and guarantee flow stability.
%whenever $P_\mathrm{dem} < \eta_\mathrm{g} \max_{K} P_\mathrm{av}(v,K)$, meaning that the power demand in a turbine is lower than maximum available power that can be extracted from the wind. 
\begin{table}[h!]
\caption{Qualitative categorization of down-regulation strategies}
\begin{center}
\begin{tabular}{c c c c}
\hline
Strategies  &  Power tracking  &  Structure Loads & Flow stability \\
\hline
Maximum $\omega_\mathrm{r}$ & High & High & Low \\
Minimum $C_\mathrm{T}$  & Low & Low & Low \\
Constant $\lambda$ & Medium & Medium & High\\
Constant $\omega_\mathrm{r}$ &  \multicolumn{1}{l}{Depending on} &  \multicolumn{1}{l}{Depending on} & Low \\
 % & \multicolumn{1}{l}{on $\omega_\mathrm{r}$} &  on $\omega_\mathrm{r}$ & \\
&  \multicolumn{1}{l}{the $\omega_\mathrm{r}$ value} &  \multicolumn{1}{l}{the $\omega_\mathrm{r}$ value} & \\
 \hline
\end{tabular}
\end{center}
\label{tablestrategies}
\end{table}

\subsection{The degree of freedom on kinetic energy}
%Considering that WTs have both generator torque and collective blade pitch angles as control input. Multiple control solutions can be found to track a desired power reference, therefore a 
% The intuitive concept of multiple control solutions and the degree of freedom on kinetic energy in down-regulation are demonstrated by considering the following proposition.

The possibility of having multiple down-regulation strategies is a consequence of the following:

\begin{proposition}\label{prop1}
There exist an non-unique steady state operating condition when a power demand is below the maximum available power. %in a constant inflow wind.
 %A constant power flow, lower than the maximum available power into the wind, can be achieved in steady state by an non-unique control solution.
\end{proposition}

\begin{figure}[h!]
\centering
\includegraphics[width=\linewidth]{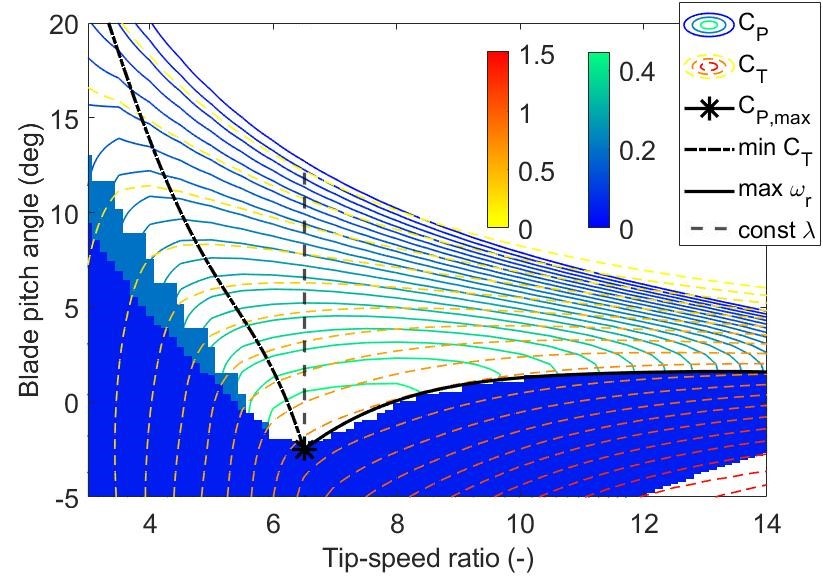}
		\caption{Operation curves of down-regulation methods and stall regions of an NREL 5MW turbine. The shades of bright and dim blue colors correspond to $\partial C_\mathrm{Q}(\lambda, \theta)/\partial \theta >0$ and $\partial C_\mathrm{Q}(\lambda, \theta)/\partial \omega_\mathrm{r}>0$, respectively, characterizing the stall regions. }
\label{allin}
\end{figure}

\begin{proof}
 Lets then assume a steady state condition with power demand and inflow wind as being $P_\mathrm{dem, \, ss}$ and  $v_\mathrm{ss}$, respectively.
%Considering keeping the kinetic energy constant and having a constant power flow, 
Also, consider the down-regulation of a turbine to be asymptotic stable by feedback~\cite{aho2012,fleming2016}, meaning that as $t \rightarrow \infty$ the demanded power flow tends to be reached by the generator, and the derivative of the kinetic energy from Eq.~\eqref{model} tends to zero in a steady state condition.
%Following \eqref{model}, the derivative of the kinetic energy will converge to zero, where a constant power flow is obtained.
Then, the following equation would hold near to an equilibrium. %the demanded power $P_\mathrm{dem}$, directly met by the generated power $P_\mathrm{g}$, is equal to the associated rotor power $\eta_\mathrm{g} P_\mathrm{r}$ from Eq. \eqref{model}, such as
\begin{equation}\label{powerss1}
0 \approx P_\mathrm{r}(t) - \frac{1}{\eta_\mathrm{g}}P_\mathrm{dem, \, ss}(t),
\end{equation}
From combining \eqref{powerss1} with \eqref{rotorpower} 
\begin{equation} \label{powerss2}
    \frac{1}{\eta_\mathrm{g}}P_\mathrm{dem, \, ss} \approx 0.5 \rho A_\mathrm{r} v_\mathrm{ss}^3 C_\mathrm{P}\left(\dfrac{\omega_\mathrm{r}(t)R}{v_\mathrm{ss}}, \theta(t)\right).
\end{equation} 

Therefore, as depicted by the $C_\mathrm{P}$ contours in Fig~\ref{allin},  a desired $C_\mathrm{P}$ value lower than the maximum $C_\mathrm{P}$ can be reached by different combinations of $\lambda$ and $\theta$.
% This leads to a freedom on the decision of the generator torque and blade pitch commands.
\end{proof}

\begin{remark}\label{consec1}
The fact that different combinations of $\lambda$ are possible means that there is an extra degree of freedom on choosing a desired rotational speed and, thus, a kinetic energy. This will indeed be leveraged to reduce aerodynamic loads and the risk of stall conditions.
\end{remark}

%\begin{figure}[t!]
%\centering
%\includegraphics[width=0.95\linewidth]{Figures/CpContourNREL2.pdf}
%		\caption{Power coefficient $C_\mathrm{P}$ curves of an NREL 5MW turbine.}
%\label{Cptable}
%\end{figure}

\subsection{Flow stability} \label{flowstab}
%The flow stability can be characterized by operation condition and observed on the steady state response.

% As a current problem on down-regulating turbines, the loss in flow stability depends on the operating condition of the turbine. The flow along the blades can be predominantly separated characterizing a stall behavior. This behavior can occur in low and high tip speed ratios and is usually problematic for the turbine control. The turbine response behaves differently than expected from changes on the generator torque and blade pitch commands. %Usually, when the generator torque or the blade pitch angles increase, the rotor speed decreases. However, as result of stall, the turbine assumes an opposite behavior.
% In addition to the challenge in control, undesirable oscillations on the turbine response is provoked by the flow separation due to induced vortexes \cite{heinz2016}. 

A current problem occurring when down-regulating turbines is due to the risk of loss of flow stability along the blades. This phenomenon, which can occur in low and high tip speed ratios, can lead to undesirable oscillations \cite{heinz2016} and stall and is thus problematic for the design of turbine control.

Stall is characterized by the decrease of lift force on turbine blades as function of their angle of attack \cite{manwell2010}. When the turbine is operating in a region where the derivatives of the aerodynamic torque with respect to rotor speed and pitch angle are positive, then it will eventually reach stall conditions~\cite{novak1995, deshpande2012, choudhry2016}. These regions are shown by the blue shades in Fig. \ref{allin} when the partial derivatives of the torque coefficient $C_\mathrm{Q}$ are positive, thus indicating the possible onset of stall.

%Therefore, the linearized constraints are extracted from the partial derivatives of torque coefficient $C_Q$ table (Fig. \ref{}), being $C_Q = C_P / \lambda$, to avoid the turbine operation under the stall region.

In particular, each down-regulation method can be analyzed in terms of flow stability from their operation distance with respect to stall regions. First, the constant $\omega_r$ strategy can easily reach stall regions with low or high tip speed ratios depending on wind speed. From the same point of view, the maximum $\omega_r$ strategy operates on the boundary of the stall pitch region.  %Although non-convex the regions can be approximated to be convex by piece-wise linear functions. Therefore, a hard constraint can be implemented relating $P_\mathrm{r}$ and $K$.
The minimum $C_\mathrm{T}$ strategy is categorized as low flow stability as its operation is close to stall regions at low tip-speed ratios.
Finally, the constant $\lambda$ always remains far from stall regions, being the strategy that operates under the most stable flow conditions.

\begin{remark} \label{rem2}
The flow stability analysis herein is based on a steady state model. Dynamic flow effects and model-plant mismatches introduce a considerable uncertainty on the stall conditions. Conservative stall constraints should therefore be considered to avoid such regions.
\end{remark}

%\begin{figure}[t!]
%\centering
%\includegraphics[width=\linewidth]{Figures/ultimate_result_const_Power2.pdf}
%		\caption{Power tracking through different down-regulation strategies at mean wind speed of 8 m/s and power reference of 1 MW. Note that for the equivalent power tracking the generator speed reaches different steady state values.}
%\label{multsolutions}
%\end{figure}

\subsection{The use of kinetic energy}
On one hand, the high kinetic energy is beneficial for power tracking, for instance, in the case where the demanded power exceeds the current maximum available power in the wind. 
In that case, the stored kinetic energy on the rotor speed is released so power tracking can be maintained longer. %Beyond that the kinetic energy can be also forced to be temporally explored as in \cite{odgaard2016}, however not included in our work.
On the other hand, high rotor speeds may lead to operation conditions close to stall regions and to higher aerodynamic loads as seen by the max $\omega_\mathrm{r}$ curve and the $C_\mathrm{T}$ contours in Fig.~\ref{allin}. In this regard, the different down-regulation strategies are further explored in Section~\ref{simul} in the convex model predictive control framework.

\section{CONVEX MODEL PREDICTIVE CONTROL} \label{CMPC}
 
 Convex MPC is based on solving a convex optimization problem and is a supported by a fairly complete body of research. Convex MPC can be solved numerically very efficiently, making it suitable to several applications.
 
%  by least-squares and linear programs.  It is well known that least-squares and linear programming problems have a fairly complete theory, arise in a variety of applications, and can be solved numerically very efficiently.
%  %solve a convex optimization problem to find the optimal control input (i.e. $P_\mathrm{g}$ and $P_\mathrm{r}$) that satisfy a multi-objective problem, which also includes an energy storage device. 

 The down-regulation in wind turbines is here formulated as an optimization problem based on the linear dynamics and convex constraints defined in Section \ref{wtmodel}. Different from previous works, such as \cite{hovgaard2013}, the turbine herein is set to track a demanded power instead of maximizing power extraction. As consequence of Proposition \ref{prop1}, a down-regulation operation is not uniquely defined, so an extra objective is added corresponding to the chosen down-regulation methodology. %This extra objective characterizes the the down-regulation strategy. 

 First, we define the extra objectives - in Table~\ref{equivalences}- and corresponding additional constraints in terms of energy and power flows. Then, further the flow stability constrain is derived. In the end, the general optimization problem is defined for all down-regulation strategies.
 
 \begin{table}[h!]
	\caption{Down-regulation strategies and equivalent objectives}
	\label{equivalences}
	\begin{center}
		\begin{tabular}{c c c}
			\hline
			Down-regulation  & Equivalent  & Weights for \eqref{costfunction}     \\
			Strategy & Objective &  [$\alpha_5$, $\alpha_6$, $\alpha_7$]  \\
			\hline
			 Maximum $\omega_\mathrm{r}$ & Maximizing $K(t)$ & [$\alpha_5$, 0, 0]  \\
			 Minimum $C_\mathrm{T}$ & Minimizing $F_\mathrm{T}(t)$ from \eqref{thrustlinear} %$\equiv$ Minimizing $R_{\mathrm{F}_\mathrm{T}} K$
			 & [0, 0, $\alpha_7$] 
			\\
		     Constant $\lambda^\mathrm{opt}$ & Tracking $K^\mathrm{ref} (t)$ & [0, $\alpha_6$, 0] \\
			 Constant $\omega_r$ & Tracking constant $K^\mathrm{ref}$ & [0, $\alpha_6$, 0] \\
			\hline
		\end{tabular}
	\end{center}
\end{table}

 %in terms of the kinetic energy associated to the rotating parts. 
%  In \cite{hovgaard2013} an energy storage is also considered, instead the energy storage is not included besides of the own kinetic energy of the turbine rotor, nor specific constraints on the power gradient.
%The extra objectives corresponding to each down-regulation strategies are presented.

The minimum $C_\mathrm{T}$ strategy is equivalent to minimize the thrust force $F_\mathrm{T}(t)$, which is also equivalent to minimize the term $R_{\mathrm{F}_\mathrm{T}} K$ from Eq.~\eqref{thrustlinear} while the rotor power $P_\mathrm{r}$ would match its associated power reference.
$R_{\mathrm{F}_\mathrm{T}}$ is a time-varying parameter, which depends on the current operation point, so an extra variable $F_\mathrm{T, \, extra}$ is instead minimized by including the following constraints.
\begin{equation} \label{minCtconst}
 (R_{\mathrm{F}_\mathrm{T}} K)' R_{\mathrm{F}_\mathrm{T}} K \leq F_\mathrm{T, \, extra},
\end{equation}
\begin{equation} \label{minCtconst}
  F_\mathrm{T, \, extra} \geq 0,
\end{equation}
where the term  $R_{\mathrm{F}_\mathrm{T}} K$ is therefore indirectly minimized based on the robust linear program \cite{boyd2004}. This is done because the term $R_{\mathrm{F}_\mathrm{T}} K$ is composed by a time-varying parameter, instead of a constant, and a decision variable.

To obtain the constant $\lambda^\mathrm{opt}$ strategy, the kinetic energy is set to track the following reference as an objective. 
\begin{equation}
    K^\mathrm{ref} (t) = \frac{J(\omega_\mathrm{g}^{\mathrm{ref}}(t))^2}{2},
\end{equation}
where
\begin{equation}
\omega_\mathrm{g}^\mathrm{ref}(t)=\dfrac{\lambda^\mathrm{opt} v(t)}{R} G_\mathrm{B}.
\end{equation}

Now, a time-varying inequality that includes a positive tuning parameter $\delta$ is introduced to constrain the turbine operation out of the stall region as
\begin{equation} \label{stallineq}
    \frac{\partial T_\mathrm{r}(t)}{\partial \theta(t)}  \leq - \delta,
\end{equation}
where, using \eqref{taeroc}, 
\begin{equation}
   \frac{\partial T_\mathrm{r}(t)}{\partial \theta(t)}  =  0.5 \rho A_\mathrm{r} R v^2(t) \dfrac{\partial C_\mathrm{Q}(\lambda(t), \theta(t))}{\partial \theta(t)}.
\end{equation}
The value of $\partial T_\mathrm{r}(t) / \partial \theta(t)$ expresses how safe the current WT operation is from stall conditions and it should be always negative as previously discussed in Subsection \ref{flowstab}. The parameter
$\delta > 0$ is recommended to be added to increase robustness as result of Remark \ref{rem2}, therefore conservatively preventing stall to happen.

Similar with the derivation of \eqref{thrustlinear}, the affine time-varying  constraint that avoids stall regions can be obtained by the linarization of the power coefficient $C_\mathrm{P}(K(t), \theta(t))$ and the derivative of the aerodynamic torque coefficient with respect to the blade pitch $\partial C_\mathrm{Q}(K(t), \theta(t))/\partial \theta(t)$. Such that the following constraint is obtained.
\begin{equation} \label{stallconst}
\frac{\partial T_\mathrm{r}(t)}{\partial \theta(t)}  \approx Q_{T_r'}(t) P_r(t) + R_{T_r'}(t) K(t) + S_{T_r'}(t) \leq - \delta,
\end{equation}
where $Q_{T_r'}(t)$, $R_{T_r'}(t)$ and $S_{T_r'}$ are the corresponding time-varying parameters. 
An affine function is always convex, so also is this constraint.

Finally, the cost function is defined as the integral of the objective function $F$ over the time horizon $T$ while  considering the linear dynamic model from \eqref{model} and the defined convex constraints from \eqref{constLinear1}, \eqref{constLinear2}, \eqref{constLinear3} and \eqref{stallconst} over the receding horizon.
\begin{equation} \label{optprob}
\textrm{max}_{U(t)} \int_{t}^{T+t} F(x(\tau),u(\tau))d\tau, \, \forall t \geq 0,
\end{equation}
\begin{equation*}
\textrm{s.t.} \, \, \eqref{model}, \eqref{constLinear1}, \eqref{constLinear2}, \eqref{constLinear3}, \eqref{stallconst},
\end{equation*}
where the objective function $F$ is defined as
\begin{equation*}
% F(x(t),u(t),d(t)) =
F(x(t),u(t)) =  
\end{equation*}
\begin{equation*}
- \alpha_1 \left[P_\mathrm{g}(t) -P^{\mathrm{ref}}(t)\right]^2 -\alpha_2 \dot{P}_\mathrm{g}^2(t)
-\alpha_3 \dot{P}_\mathrm{r}^2(t) 
\end{equation*}
\begin{equation*}
-\alpha_4 \left[ \max  \{ K(t) - \frac{J}{2} \omega_\mathrm{g,rated}^2 , \, 0 \} \right] +\alpha_5 K(t)
\end{equation*}
\begin{equation} \label{costfunction}
-\alpha_6 \left[ K(t) - K^\mathrm{ref}(t) \right]^2    -\alpha_7 F_\mathrm{T, \, extra}(t).
\end{equation}

\noindent In the previous equations, it holds $x(t) = [K(t)] $, $u(t) = [P_\mathrm{r}(t), P_\mathrm{g}(t)]^\top$,  and $U(t) = [u^\top(t), ..., u^\top(t+T)]^\top$.
For each down-regulation strategy, the corresponding weights for $F$ are listed in Table \ref{equivalences}, where $\alpha_5$, $\alpha_6$, $\alpha_7 \in \mathbb R_{> 0}$. The additional constraint \eqref{minCtconst} for the minimum $C_\mathrm{T}$ strategy is set only when this strategy is chosen.
%\begin{equation}
%-\alpha_9 \left[ F_T(t) -F_T(t-1) \right]^2
%\end{equation}
%where the terms associated with $\alpha_5$, $\alpha_6$, and $\alpha_7$ defines the extra objectives given the degree of freedom.
\begin{remark} \label{rem3}
Stability is usually guaranteed by including a terminal cost and terminal constraints. However, the formulation in \eqref{optprob} does not include them. Therefore, we are aware that this choice does not yield a closed-loop
stability guarantee.
\end{remark}

In the proposed formulation, the optimization problem can be solved globally using efficient algorithms~\cite{boyd2004}. At a defined sampling time $T_s$, the optimal solution for $U(t)$ as a vector sequence along the time horizon $T$ is obtained whereby the first input of the sequence is used to compute the current turbine command. Then, the prediction horizon moves ahead to the next step time, from which the optimization is repeated.  

In particular, given the values of $P_\mathrm{r}$, $P_\mathrm{g}$ obtained from the MPC solution, we can obtain the blade pitch command  
from Eq.~\eqref{rotorpower} as 
\begin{equation} \label{pitchcommand}
    \theta(P_\mathrm{r}(t), K(t)) = \theta_\mathrm{table} \left( \frac{P_\mathrm{r}(t)}{0.5 \rho A_\mathrm{r} v^3(t)},  \frac{R\sqrt{(2/J) K(t)}}{v(t)} \right),
\end{equation}
and the generator torque command from Eq.~\eqref{powergen} as
\begin{equation} \label{torqcommand}
    T_\mathrm{g}(P_\mathrm{g}(t),K(t)) = \frac{P_\mathrm{g}(t)}{\eta_\mathrm{g}\sqrt{(2/J) K(t)}}.
\end{equation}

\section{SIMULATIONS} \label{simul}

In this section, the performances of the controller with different down-regulation strategies described in the previous section are compared for the case where a time-varying demanded power exceeds the current maximum available power in the wind. The control parameters used in the OpenFAST simulations are in Table \ref{controlpar}. The normalized standard test signal from  \cite{pilong2013} is used for the reference power as in \cite{fleming2016}, where the time-varying reference power signal is herein set to exceed the maximum available power. The maximum available power is derived from the simulated constant and uniform wind speed of 8 m/s, that is a representative average annual value in a wind plant site.

\begin{table}[h!]
	\caption{Control parameters}
	\label{controlpar}
	\begin{center}
		\begin{tabular}{c c}
			\hline
			Parameter  & Value    \\
			\hline
			Weights, $\alpha_i$  &  [10 0.1 0.1 0.1 0.1 0.1 0.01]\\
			Time horizon, $T$ & 20 s \\
			Sampling time, $T_s$ & 0.2 s \\
			Stall constraint parameter, $\delta$ & 0 \\
			\hline
		\end{tabular}
	\end{center}
\end{table}

 \begin{table*}[t!]
	\caption{Comparison of Controller Power Output Tracking and Previous Aerodynamic Loading}
	\label{compare}
	\begin{center}	\begin{tabular}{c c c c}
			\hline
			 Down-regulation  & Kinetic Energy (J) & Mean Load (kN) & Time of Power Tracking  \\
			 Strategy & [before 300 s] & [before 300 s] &  after Saturation (s)\\
			\hline
	     Maximizing $K(t)$ &  4.8675e+07 & 306 & 40\\
	     Minimizing $\hat{F}_\mathrm{T} (t)$ %$\equiv$ Minimizing $R_{\mathrm{F}_\mathrm{T}} K$
			& 0.7689e+06 &  259  & 4 
			\\
		  Tracking $K^\mathrm{ref}(t)$ & 1.3750e+07 & 264 & 5 \\
		 Tracking constant $K^\mathrm{ref}$ &  1.5086e+07 & 266 & 9\\
			\hline
	\end{tabular}
	\end{center}
\end{table*}

For each down-regulation strategies, the mean values of kinetic energy and aerodynamic loads are computed before the power reference increases at 300 s. The amount of time that turbines can track the required power after it becomes saturated are presented in Table \ref{compare}.
\begin{figure}[b!]
\centering
\includegraphics[width=\linewidth]{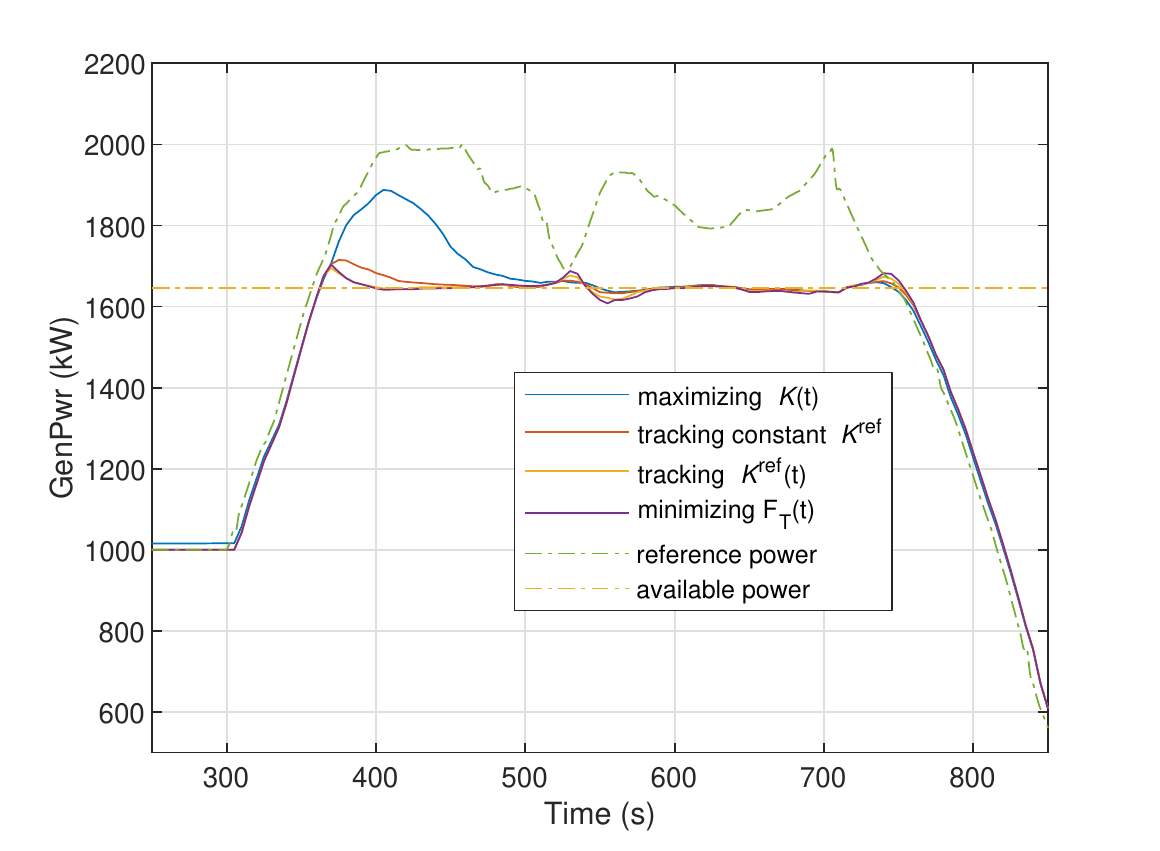}
	\caption{Active power signals (NREL 5MW).}
\label{powervar}
\end{figure}
The turbine's power and reference are depicted in Fig.~\ref{powervar}. The reference power in green is set to exceed the available power in orange. The generated powers from each down-regulation strategy present overshoots with respect to the available power when saturation occurs due to the stored kinetic energy. The strategy that maximizes kinetic energy follows longer the reference power, although the turbine operates under higher aerodynamic loads.

%compares the RMS error of the tracking after turbine saturation, and kinetic energy and aerodynamic loads before the increase of power demand across the methods.

%\begin{figure}[h]
%\centering
%\includegraphics[width=\linewidth]{Figures/comparation2b.pdf}
%	\caption{Turbine state and inputs (NREL 5MW).}
%\label{turbstates}
%\end{figure}

%\subsection{Pareto front}
%Depending on the designed power demanded for a specific turbine and average wind conditions, a Pareto front \cite{odgaard2016b} can be obtained once two of the extra objectives are from interest, namely, the maximization of kinetic energy and the minimization of the aerodynamic loads, for instance. 

%The new positive parameter $\gamma<1$ for the tuning of the two extra objectives is added by multiplying (1-$\gamma$) and $\gamma$ to the corresponding weights $\alpha_5$ and $\alpha_7$, respectively, in the objective function. The middle point search method is conducted based on the gradient of the performance indicators.

\section{CONCLUSIONS} \label{conc}

% In this paper, several strategies for turbines down-regulation are implemented via convex model predictive control. By tracking a requested power, a non-unique solution in term of rotor speed - translated to a degree of freedom on the kinetic energy - allows turbines to pursue an extra objective. This objective defines the down-regulation strategy. Also, a constraint for the flow stability is derived to avoid vortex-induced loading. We have demonstrated the results of different down-regulation methodologies in terms of structural loading, flow stability and power tracking capability. 
% Simulation on realistic models reveal the ability to maintain power tracking in saturation conditions by taking advantage of the kinetic energy of the rotating parts.

% %This work demonstrates the flexibility on controlling wind turbines within appropriated constraints and the potential on the use of kinetic energy

% Facing the challenge of variability, uncertainty and asynchronism on wind energy, active power control of wind turbines can benefit from making use of the turbine's kinetic energy to full-fill the future grid code requirements being a research target as in \cite{aho2012, ela2014}. Although the economic impact is still the aim of current research, the shifting paradigm from maximizing to tracking power and the use of kinetic reserve present a significant economic potential through the development of wind turbine controllers.

In this paper, we proposed a linear convex model predictive control framework for implementing wind turbine down-regulation. Down-regulating a turbine leads to an additional degree of freedom on the value of the kinetic energy. We leveraged this and shown how different choices of the target kinetic energy can lead to existing down-regulation strategies. We further introduced a novel strategy aimed at reducing aerodynamic loads and reducing the risk of stall conditions. We have demonstrated the results of different down-regulation methodologies in terms of structural loading, flow stability and power tracking capability. 
Simulation on realistic models reveal the ability to maintain power tracking in saturation conditions, by means of the stored kinetic energy of the rotor.

%This work demonstrates the flexibility on controlling wind turbines within appropriated constraints and the potential on the use of kinetic energy

The shifting paradigm from maximizing to tracking power and the use of kinetic energy as a storage  presents a significant economic potential and encourages the research on active power control of wind turbines \cite{aho2012, ela2014}. As future work, the extension of the proposed MPC approach for the problem of power dispatch in wind farm control, and the effects of wakes, is of particular interest.

                            
%%%%%%%%%%%%%%%%%%%%%%%%%%%%%%%%%%%%%%%%%%%%%%%%%%%%%%%%%%%%%%%%%%%%%%%%%%%%%%%%
%\section*{APPENDIX}

\section*{ACKNOWLEDGMENT}

The authors acknowledge support from the European Union Horizon 2020 program through the WATEREYE project (grant no. 851207). 
The authors would like also to thank Atindriyo Pamososuryo from TU-Delft for sharing the baseline simulation code. 
%The preferred spelling of the word  acknowledgment  in America is without an  e  after the  g . Avoid the stilted expression,  One of us (R. B. G.) thanks . . .   Instead, try  R. B. G. thanks . Put sponsor acknowledgments in the unnumbered footnote on the first page.

%%%%%%%%%%%%%%%%%%%%%%%%%%%%%%%%%%%%%%%%%%%%%%%%%%%%%%%%%%%%%%%%%%%%%%%%%%%%%%%%
\bibliographystyle{IEEEtran}
\bibliography{IEEEabrv,IEEEexample}

\end{document}